# HIDDEN PERCOLATION TRANSITION IN KINETIC REPLICATION PROCESS

P. N. Timonin[1], G. Y. Chitov[2]

*(1) Physics Research Institute, Southern Federal University,*

*344090, Stachki 194, Rostov-on-Don, Russia, pntim@live.ru*

*(2) Department of Physics, Laurentian University*

*Sudbury, Ontario P3E 2C6, Canada, gchitov@laurentian.ca*

*The one-dimensional kinetic contact process with parallel update is introduced and studied by the mean-field approximation and Monte Carlo (MC) simulations. Contrary to a more conventional scenario with single active phase for 1d models with Ising-like variables, we find two different adjacent active phases in the parameter space of the proposed model with a second-order transition between them and a multiphase point where the active and the absorbing phases meet. While one of the active phases is quite standard with a smooth average filling of the space-time lattice, the second active phase demonstrates a very subtle (hidden) percolating order which becomes manifest only after certain transformation from the original model. We determine the percolation order parameter for active-active phase transition and discuss such hidden orders in other low-dimensional systems. Our MC data demonstrate finite-size critical and near-critical scaling of the order parameter relaxation for the two phase transitions. We find three independent critical indices for them and conclude that they both belong to the directed percolation universality class.*

# 1. Introduction

There has been a steadily growing interest during the last two decades or so in the kinetics and phase transitions in non-equilibrium systems. The applications of those systems range from physics, like, e.g., statistical physics, critical phenomena, condensed matter to less conventional fields, like biology, ecology or quantitative finance [1- 5]. The central problem in studies of non-equilibrium systems is the transitions they undergo between various active phases and the inactive (absorbing) state. The kinetic contact processes which model, e.g., the epidemic disease spreading or population replication, are the simplest kinetic models exhibiting such non-equilibrium phase transitions under variation of their parameters [1- 5].

In the present paper we propose and study a one-dimensional kinetic contact process with parallel update (probabilistic cellular automaton, PCA). Along with the trivial absorbing state, we find two active phases. The examples of multiple active phases and transitions between them are actively discussed in the current literature in case of models in spatial dimensions $d \geq 2$ [6-10] and in $1d$ models with $n$-valued ($n > 2$) variables [11-14]. However, the short-range models with Ising-like (two-valued) variables which have a single absorbing state and no conservation laws are known to exhibit only one transition into a single active phase belonging to the directed percolation (DP) universality class [1]. The present study shows that such models can possess two or more active phases since what is now believed to be a single active phase can be actually several ones with transitions between them. To reveal such hidden phases and transitions a close inspection of steady state patterns may be needed as the emergent long-range order can look like gradual changes of a short-range one. Here we show that it is the case for the proposed model. In a certain range of the model parameters the crucial change of the short-range order emerges through the appearance of large regions with disbalance of sublattice occupations ("antiferromagnetic domains") in the active phase. A naïve mean-field treatment predicts the additional "antiferromagnetic" (AF) order in this region, but the situation in fact is more subtle: if we do not discriminate the two types of the antiferromagnetic clusters and merge adjacent ones we find that such AF order percolates infinitely in the time direction. To demonstrate such a percolation one needs to perform certain transformations of the lattice and occupation number variables of the original model. We compute the order parameter for the transition to the "antiferromagnetic percolation" phase as the capacity of percolation cluster on the constructed dual decorated square lattice where the AF order percolates. The rest of the active region of parameter space where the AF percolation is absent constitutes the other "ferromagnetic" phase, which has a conventional order corresponding to a uniform in average filling of the sites.

The paper is organized as follows: In the next Section 2 we introduce and discuss the PCA model to study. In Sec. 3 we present the results of the mean-field analysis of the model. Sec. 4 contains the direct numerical results for the phase diagram and critical behavior of the present PCA. In Sec. 5 we give an alternative statistical-mechanical formulation of the model and analytical representation for the nonlocal (string) order parameter to deal with the percolation. The results are summarized in the last Sec. 6.

## 2. Model

In this paper we introduce a one-dimensional PCA representing the contact process of population replication or disease spreading. Imagine a line of plants which maintain population by spreading the seeds to the nearby sites so the new plants can grow on them if these sites are empty. Let $q$ be the probability of such event for empty site having only one neighboring plant. For the case of two plants around the empty one we can choose this probability $r > q$ to be $r = 1-(1-q)^2 = q(2-q)$ assuming the independence of two plants' seeds "not spreading effects". Some other choices are also possible but here we consider only the case $r = q(2-q)$. Let also $p$ be the probability for plant to survive in the one time step. Thus we have the PCA on the line of two-state sites (empty and full, $n_i = 0, 1$) with the evolution step probabilities defined for the local three-site configurations in Table 1. The objects $n_i$ residing on the sites can be thought of as classical particles, analogous to hard-core bosons in the classical limit.

Table 1.

| $n_{i-1}, n_i, n_{i+1}$ | 111, 110, 011, 010 | 100, 001 | 101 | 000 |
|---|---|---|---|---|
| Probability ($n_i = 1$) | $p$ | $q$ | $q(2-q)$ | 0 |

At $p = 0$ the present model is equivalent to the bond directed percolation (BDP) [1]. Indeed, if in the initial state the particles are present on the even (or odd) sites only, then the evolution is of the BDP-type. Thus on the $2d$ time-space lattice just one sublattice, say A (cf. Fig. 1(a)), becomes partially filled, while the other sublattice B stays empty. At $q > q_{BDP} \approx 0.6447$ [1] a percolation cluster is formed on the one sublattice, see Fig. 1(b). This order which we could call "antiferromagnetic" (AF) is still preserved in series of "antiferromagnetic domains" when starting with random initial state as shown in Fig. 1(c). The dotted lines in Fig. 1(c) separate the regions where the particles exist on one sublattice only. The percolation cluster is formed at $q > q_{BDP}$ as our Monte Carlo data show (see below).

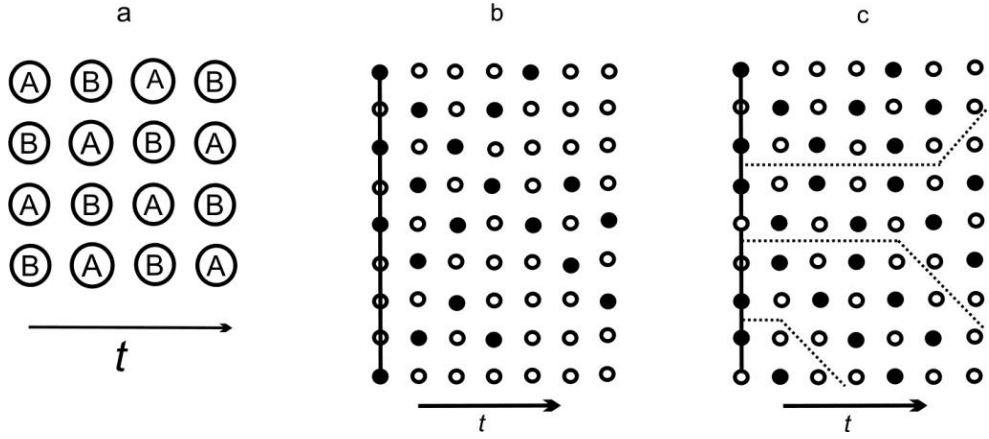

Fig.1. *Evolution in the replication process at p = 0. The filled/empty circles correspond to n=1 or 0, resp. (a) Division of the 2d lattice on two sublattices A and B; (b) Evolution of the initial state with subjects on even (odd) lattice sites only; (c) Evolution of the random initial state, dotted lines are the boundaries between the "antiferromagnetic domains".*

This behavior could be contrasted with that at $p = 1$, $q \neq 0$ when the limiting stable state is a fully occupied time-space lattice. We call this state "ferromagnetic" (F). So during the evolution of $p$ from 0 to 1 the stable state of the PCA undergoes a drastic change from the "antiferromagnetic" to the "ferromagnetic" order, so we can assume that there is a phase transition between these two active phases. In this paper we explore the possibility of such transition in the present model using both the mean-field and Monte-Carlo techniques.

### 3. Mean-field approximation

The state of the system at the time step $t$ is specified by the occupation numbers $n_{t,i} = 0,1$, $i = 0,1,...,N-1$. The process is defined via the conditional probabilities

$$W(\mathbf{n}_{t+1}|\mathbf{n}_t) = \prod_{i=0}^{N-1} W(n_{t+1,i}|n_{t,i-1}, n_{t,i}, n_{t,i+1}) \qquad (1)$$

which yield the probabilities of the states after $T$ time steps as

$$W(\mathbf{n}_T|\mathbf{n}_0) = \sum_{\mathbf{n}_1,...,\mathbf{n}_{T-1}} \prod_{t=0}^{T-1} W(\mathbf{n}_{t+1}|\mathbf{n}_t) \qquad (2)$$

For the present PCA

$$W(n_{t+1,i}|n_{t,i-1}, n_{t,i}, n_{t,i+1}) = \{n_{t+1,i} P_{t,i}(\mathbf{n}_t) + (1 - n_{t+1,i})[1 - P_{t,i}(\mathbf{n}_t)]\} \qquad (3)$$

$$P_{t,i}(\mathbf{n}_t) = p n_{t,i} + (1 - n_{t,i})\{q[n_{t,i-1}(1 - n_{t,i+1}) + n_{t,i+1}(1 - n_{t,i-1})] + q(2-q) n_{t,i-1} n_{t,i+1}\} \qquad (4)$$

The evolution of any initial distribution function of the states is governed by the equations

$$\rho_{t+1}(\mathbf{n}_{t+1}) = \sum_t W(\mathbf{n}_{t+1}|\mathbf{n}_t)\rho_t(\mathbf{n}_t) \tag{5}$$

Since $\sum_{n_{t+1,i}} W(n_{t+1,i}|n_{t,i-1},n_{t,i},n_{t,i+1}) = 1$, we can sum Eq. (5) over all $n_{t+1,i}$ except one, say, $n_{t+1,1} \equiv n_1$, to get the evolution equations for the partial distribution functions

$$\rho_{t+1}^{(1)}(n_1) = \sum_{n_0',n_1',n_2'} W(n_1|n_0',n_1',n_2')\rho_t^{(3)}(n_0',n_1',n_2') \tag{6}$$

The simplest mean-field approximation consists in decoupling of the three-particle distribution function into the single-particle functions as

$$\rho_t^{(3)}(n_0',n_1',n_2') = \rho_t^{(1)}(n_0')\rho_t^{(1)}(n_1')\rho_t^{(1)}(n_2')$$

thus giving a closed equation for the single-particle distribution function

$$\rho_{t+1}^{(1)}(n_1) = \sum_{n_0',n_1',n_2'} W(n_1|n_0',n_1',n_2')\rho_t^{(1)}(n_0')\rho_t^{(1)}(n_1')\rho_t^{(1)}(n_2') \tag{7}$$

Anticipating an AF order with different occupations $n_{t,i}$ on the sublattices A with $t+i=2k$ and B with $t+i=2k+1$, we introduce two different one-particle distribution functions $\rho_t^A(n)$ and $\rho_t^B(n)$ for each sublattice. Then Eq. (7) applied for different sublattices yields

$$\rho_{t+1}^{(A)}(n_1) = \sum_{n_0',n_1',n_2'} W(n_1|n_0',n_1',n_2')\rho_t^{(A)}(n_0')\rho_t^{(B)}(n_1')\rho_t^{(A)}(n_2') \tag{8}$$

$$\rho_{t+1}^{(B)}(n_1) = \sum_{n_0',n_1',n_2'} W(n_1|n_0',n_1',n_2')\rho_t^{(B)}(n_0')\rho_t^{(A)}(n_1')\rho_t^{(B)}(n_2') \tag{9}$$

These equations preserve normalization of $\rho_t^{(A,B)}(n)$, i.e, $\rho_t^{(A,B)}(0) + \rho_t^{(A,B)}(1) = 1$, so they give only two independent equations for $\rho_t^{(A,B)}(1) \equiv \varphi_t^{(A,B)}$:

$$\varphi_{t+1}^A = p\varphi_t^B + q\varphi_t^A(2-q\varphi_t^A)(1-\varphi_t^B)$$
$$\varphi_{t+1}^B = p\varphi_t^A + q\varphi_t^B(2-q\varphi_t^B)(1-\varphi_t^A) \tag{10}$$

Eqs. (10) have a trivial stationary point $\varphi^A = \varphi^B = 0$ which is stable at $p < 1 - 2q$. So this region on the ($p$, $q$) plane corresponds to the absorbing phase (A), see Fig. 2. The other stationary point of Eq. (10) is

$$\varphi^A = \varphi^B = \frac{2(2q-1+p)}{q\left(2+q+\sqrt{(2-q)^2+4(1-p)}\right)}, \quad q \neq 0 \tag{11}$$

It exists at $p > 1 - 2q$ and is stable if

$$p > p_c(q) = \frac{2-q}{8}\left(3q+2-\sqrt{9q^2+20(1-q)}\right) \tag{12}$$

Note also that at $q = 0$ and $p = 1$ we have a deterministic evolution in which starting values $\varphi_0^{(A)}$ and $\varphi_0^{(B)}$ interchange at each time step, such that $\varphi_t^{(A)} + \varphi_t^{(B)} = \varphi_0^{(A)} + \varphi_0^{(B)}$.

Thus at $p > \max(p_c(q), 1-2q)$ the model evolves into the "ferromagnetic" (F) stationary phase with equal occupations on the sublattices A and B, see Fig.2. In the third region of the phase diagram where $p < p_c(q)$ and $q > 1/2$ we find a "ferrimagnetic" (F+AF) phase with stable stationary points having different sublattice occupations:

$$\varphi^A + \varphi^B = \frac{(2-q)(2q-1-p)}{q(2q-q^2-p)} \tag{13}$$

$$\varphi^A - \varphi^B = \pm \frac{\sqrt{(2q-1-p)\left[(2q-1)(2-q)^2 - (2+3q)(2-q)p + 4p^2\right]}}{q(2q-q^2-p)} \tag{14}$$

Here $\varphi^A + \varphi^B$ has a meaning of the "ferromagnetic" order parameter giving the average site occupation, while $\varphi^A - \varphi^B$ can be interpreted as an "antiferromagnetic" order parameter measuring the difference of the sublattice occupations.

The resulting phase diagram is shown in Fig. 2 (dotted lines). All transitions between the phases are of the second order. According to Eq. (11) the critical index of the order parameter for the A – F transition is $\beta = 1$, while $\beta = \frac{1}{2}$ for the F – (F+AF) transition, cf. Eq. (14). The behavior of $\varphi^A$ and $\varphi^B$ as functions of $p$ is shown in Fig. 3 for $q = 0.8$ and $q = 1$. Notice that at $p = 0$ $\varphi^A = (2q-1)/q^2$, $\varphi^B = 0$ (or vice versa) so just one sublattice is filled while the other is empty, thus the mean field predicts a pure AF state depicted in Fig. 1(a).

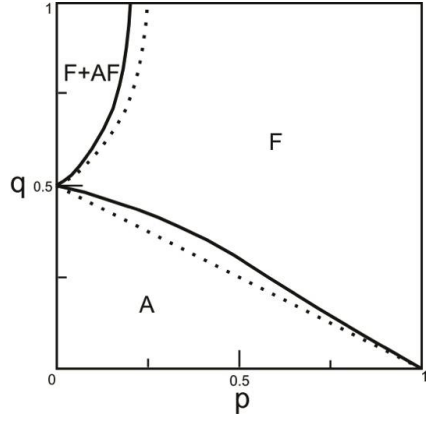

Fig. 2. *The mean-field phase diagram. The dotted and solid lines are the phase boundaries in the one-particle and the two-particle approximations, respectively.*

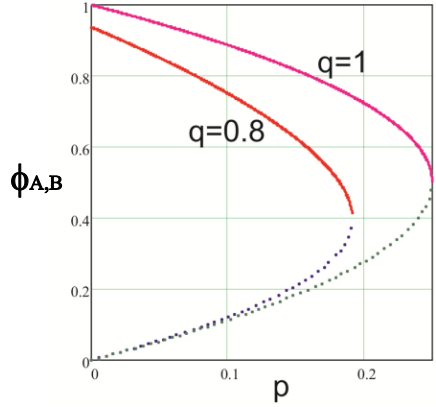

Fig. 3. *(Color online) The average occupation of the sublattices A (solid lines) and B (dotted lines) in the F+AF phase at q = 0.8 and q = 1.*

We should also mention some specific features of the F–phase at $p = 1$, $q \to 0$. At $p = 1$, $q = 0$ arbitrary stationary values of $\varphi_t^{(A)} + \varphi_t^{(B)} = \varphi_0^{(A)} + \varphi_0^{(B)}$ are allowed, while for $p = 1$, $q > 0$ the solutions are $\varphi^A = \varphi^B = 1$, so it is the point of the first order A – F transition. Near it $\varphi^A = \varphi^B \equiv \varphi$ exhibits a nonanalytic behavior $\varphi \approx (2q - 1 + p)/2q$: its limiting value at $p \to 1$, $q \to 0$ depends on the direction the limit is approached. This nonanaliticity is a consequence of the change of the order of transition: from the second at $p < 1$ into the first at $p = 1$.

We can obtain an improved phase diagram from the mean-field equations for the two-particle distribution function. Using the following decoupling of the exact distribution function

$$\rho_t^{(4)}(n_0', n_1', n_2', n_3') = \rho_t^{(2)}(n_0', n_1')\rho_t^{(2)}(n_2', n_3')$$

we get

$$\rho_{t+1}^{(2)}(n_1,n_2) = \sum_{n_0',n_1',n_2',n_3'} W(n_1|n_0',n_1',n_2')W(n_2|n_1',n_2',n_3')\rho_t^{(2)}(n_0',n_1')\rho_t^{(2)}(n_2',n_3') \tag{15}$$

Again, these equations preserve normalization of $\rho_t^{(2)}$:

$$\rho_t^{(2)}(0,0) + \rho_t^{(2)}(1,0) + \rho_t^{(2)}(0,1) + \rho_t^{(2)}(1,1) = 1,$$

so there are only three independent equations in (15). The single-particle parameters are recovered from the following relations:

$$\varphi_t^A = \rho_t^{(2)}(1,1) + \rho_t^{(2)}(1,0), \quad \varphi_t^B = \rho_t^{(2)}(1,1) + \rho_t^{(2)}(0,1) \tag{16}$$

The F-phase corresponds to the stationary point with $\rho_t^{(2)}(1,0) = \rho_t^{(2)}(0,1)$, while that with $\rho_t^{(2)}(1,0) \neq \rho_t^{(2)}(0,1)$ designates the F+AF order.

We introduce three independent variables

$$\varphi_t = \rho_t^{(2)}(1,1) + \left[\rho_t^{(2)}(1,0) + \rho_t^{(2)}(0,1)\right]/2$$

$$\psi_t = \rho_t^{(2)}(0,0), \quad \lambda_t = \left[\rho_t^{(2)}(1,0) - \rho_t^{(2)}(0,1)\right]/2$$

According to Eq. (16), $\varphi_t$ is the average occupation number for the whole chain and $\lambda_t$ is the order parameter for the F–(F+AF) transition,

$$\varphi_t^A = \varphi_t + \lambda_t, \quad \varphi_t^B = \varphi_t - \lambda_t. \tag{17}$$

As follows from (15):

$$\varphi_{t+1} = q + (p-q)^2 \varphi_t + q\psi_t(q\varphi_t - 1) + q(1-q)(\lambda_t^2 - \varphi_t^2)$$

$$\psi_{t+1} = (1-q)^2(1-2p\varphi_t) + 2q(1-q)\psi_t(1-p\varphi_t) + q^2\psi_t^2 +$$
$$+ \left[p^2 + q(2-q)(1-2p)\right](\varphi_t^2 - \lambda_t^2) \tag{18}$$

$$\lambda_{t+1} = \lambda_t\left[q(2-q) - p + q^2\psi_t\right]$$

The stationary point of these equations $\varphi = \lambda = 0$, $\psi = 1$ is stable at

$$p < p_0(q) = \frac{1-2q}{1-2q^2}$$

so this region corresponds to the absorbing phase A. Another stationary point corresponds to the solution

$$\varphi = \frac{(2q-1-p)\left[q(2-q)+p(2q^2+3q-1)+p^2\right]}{2q\left[q^2(2-q)+p(q^3-5q+1)+p^2(1+3q-q^2)-p^3\right]},$$

$$\psi = \left[(1-q)^2+p\right]/q^2, \qquad (19)$$

$$\lambda^2 = \varphi^2 - \frac{p(2q-1-p)\left[(1-q)^2+q(1-p)\right]}{q^2\left[q^2(2-q)+p(q^3-5q+1)+p^2(1+3q-q^2)-p^3\right]}.$$

The stability region of this solution is $q > 1/2$, $p < p_c(q)$, and it corresponds to the (F+AF)-phase. The function $p_c(q)$ is a solution of the fifth-order polynomial equation (19) for $\lambda = 0$, so it cannot be presented analytically. In general it grows monotonously from $p_c(1/2) = 0$ to $p_c(1) = 1/5$. In the F-phase at $p > \max\left[p_c(q), p_0(q)\right]$ the stationary point with $\lambda = 0$, $\varphi \neq 0$, $\psi < 1$ is stable. The resulting phase diagram is shown in Fig. 2 by the solid lines. As one can see from Fig. 2, the two-particle mean-field approximation results in a somewhat reduced region of the (F+AF)-phase and a slightly extended region of the absorbing A–phase in comparison to the results of the single-particle mean-field approximation. Yet the qualitative features of all transitions are the same in both approximations. In particular, at $p = 0$: $\varphi^A = (2q-1)/q^2$, $\varphi^B = 0$ (or vice versa), cf. Eq.(17).

### 4. Phase diagram of the model: Numerical results

Since for low-dimensional models like the one we study here, the mean-field treatments result in mostly qualitatively suggesting predictions at best, we resort to direct numerical simulations for our model to determine its stationary properties on the parametric $(p,q)$-plane, i.e., its phase diagram.

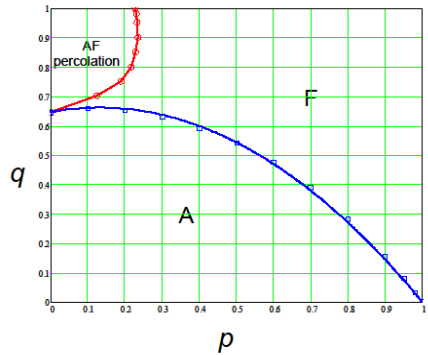

Fig. 4. *(Color online) Phase diagram from the Monte Carlo simulations. Circles and squares are the MC data, the lower blue line is the parabolic approximation Eq. (20), the upper red line represent approximate boundary between the AF percolated and the ferromagnetic phases. The both lines correspond to the continuous phase transitions.*

Our Monte Carlo (MC) studies show that the model do have second order transition between absorbing (A) and "ferromagnetic" (F) phases along the line shown in Fig. 4.

It fits approximately the following parabolic law, see Fig. 4,:

$$q = (1-p)(q_{BDP} + ap), \quad q_{BDP} \approx 0.645, \, a \approx 0.88 \qquad (20)$$

The order parameter for A-F transition is the infinite-time limit of the concentration of active sites

$$\rho(t) = \sum_{i=1}^{N} n_{t,i} / N$$

To determine the transition line and critical indices for this transition we studied near-critical and critical relaxation of $\rho(t)$ using standard protocol suggested in Ref. [1], ch. 3. We use samples with $N = 10^5$ sites and monitor $\rho(t)$ near-critical relaxation during 2000 time steps averaging these data over 1000 trials. Critical relaxation we studied on samples with $N = 100, 200, 400$ sites for $T = 5 \cdot 10^3$, $1.5 \cdot 10^3$, $5 \cdot 10^4$ time steps averaging data over $2 \cdot 10^4$, $10^4$ and $5 \cdot 10^3$ trials correspondingly. The example of such simulations establishing the validity of critical scaling is shown in Fig. 5. We get the index $\alpha$ describing critical relaxation of the order parameter, correlation length index $\nu_\parallel$ for time-direction and dynamical index $z = \nu_\parallel / \nu_\perp$. The indices obtained are presented in Table 2.

Table 2. *Critical points $q_c$ and critical indices for the transition between the absorbing (A) and ferromagnetic (F) phases for several values of p. The last row presents the known DP indices [1] for comparison.*

| p | $q_c$ | $\alpha$ | $\nu_\parallel$ | z | $\beta = \alpha \nu_\parallel$ | $\nu_\perp = \nu_\parallel / z$ |
|---|---|---|---|---|---|---|
| 0.8 | 0.2821(66) | 0.16(01) | 1.71(01) | 1.54(7) | 0.27(4) | 1.10(5) |
| 0.5 | 0.54151(05) | 0.16(01) | 1.71(4) | 1.57(6) | 0.27(4) | 1.08(8) |
| 0.2 | 0.6510(58) | 0.16(01) | 1.72(1) | 1.56(6) | 0.27(5) | 1.09(9) |
| 0 | 0.6446(6) | 0.16(01) | 1.72(5) | 1.56(4) | 0.27(6) | 1.10(3) |
| DP | 0.6447(7) | 0.159(5) | 1.73(4) | 1.58(1) | 0.27(6) | 1.09(1) |

The numerical values of the indices lie closely to the directed percolation (DP) values [1, 2] so we assume this transition belongs to the DP universality class. Note that at $q \to 0$ the PCA considered here tends to the ordinary pair contact process [1, 2], which is known to be of DP class[1]. Also, when starting from general initial states at $p = 0$ we have two BDP processes evolving on sublattices A and B respectively and our simulations show that their interference does not change the critical properties of the transition. Thus we expect the DP universality class near two extreme points of the line (20), and the data in Table 2 imply that it is true on the whole A-F line.

---

[1] We thank the anonymous referee of the paper for this observation.

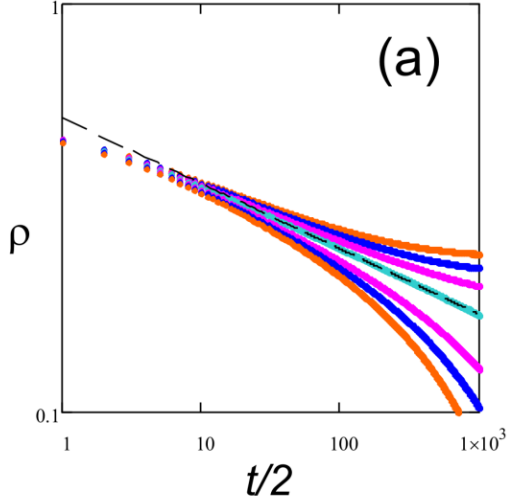
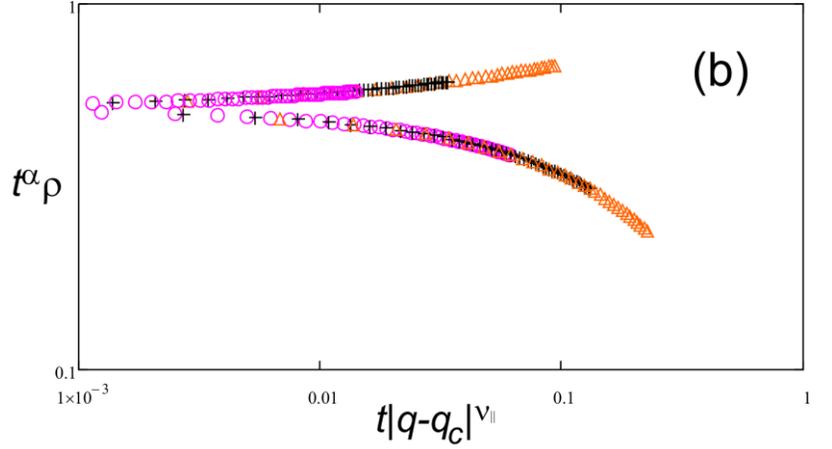
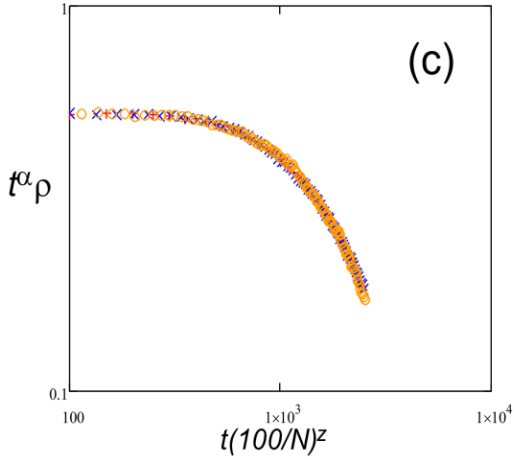

Fig. 5. *(Color online) (a) MC simulations of $\rho(t)$ relaxation for $p = 0.5$ and series of $q$ near $q_c \approx 0.542$, from top to bottom: $q=0.546, 0.544, 0.543, 0.542, 0.538, 0.536, 0.534$. Dashed line corresponds to power law $0.52/t^\alpha$, with $\alpha = 0.16$. (b) Collapse of the curves from (a) onto a single scaling function. Fitting gives the values of $\nu_{\parallel}$ and $q_c$ shown in Table 2. (c) Relaxation of the order parameter for chains with numbers of sites $N =100$ (x), $N =200$ (+), $N =400$ (o) at $q=q_c$ collapsed onto a single scaling function yields the index $z$.*

The MC data confirm the mean-field result on the first order transition into the fully occupied state at $p = 1$. In particular, in agreement with the mean-field prediction we found that $\rho(\infty)$ reaches different limiting values while approaching the point $p = 1$, $q = 0$ along different lines, see Fig. 6.

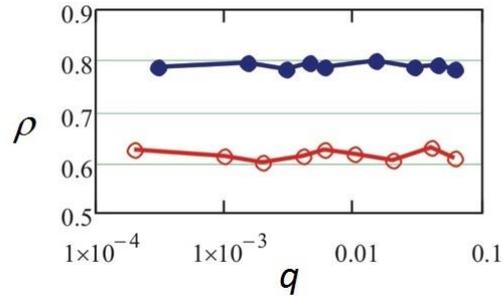

Fig. 6. *(Color online) $q$-dependence of $\rho(\infty)$ near $p = 1$, $q = 0$ along the lines $p =1-q/2$ (filled circles) and $p =1-q/3$ (empty circles).*

The MC simulations do not reveal the nonzero global order parameter $\lambda = \varphi_A - \varphi_B \neq 0$ predicted by the mean-field analysis for the (F+AF)-phase, even for a fully AF initial state. Inspecting the steady states patterns shown in Fig. 7 we conclude that mean-field prediction of the spontaneous occupation disbalance on the sublattices A and B at small $p$ is not however completely wrong, since the disbalance can be seen in some macroscopic regions of the $2d$ lattice spreading through all times probed.

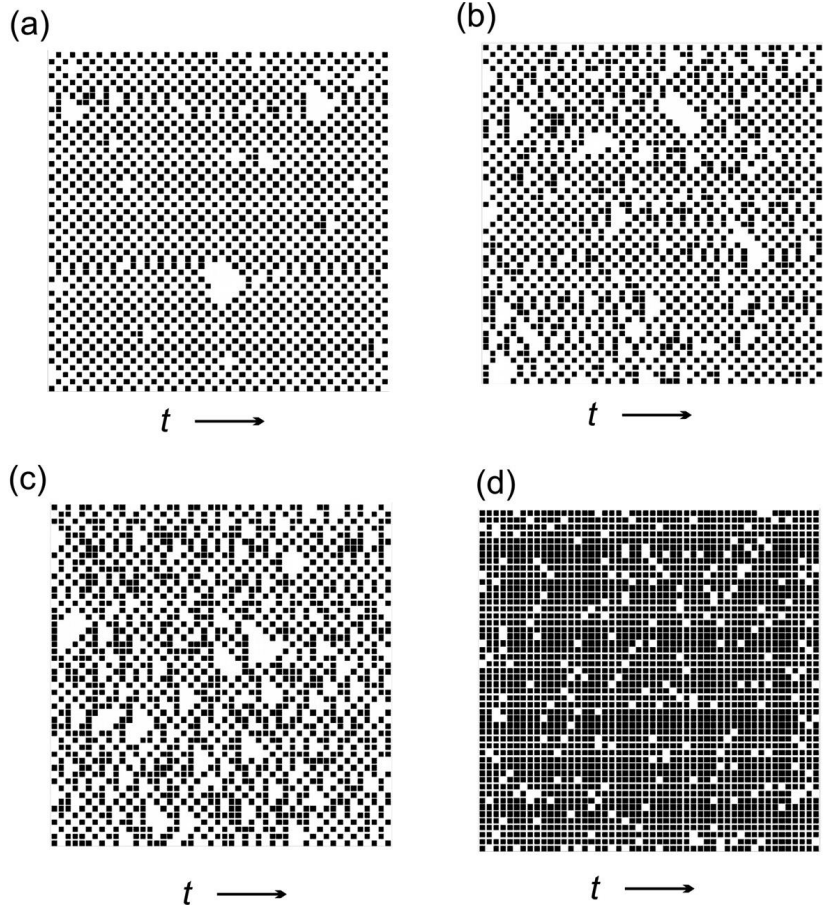

Fig. 7. *Regions of the active states stable configurations in two different active phases ($q = 0.9$). The AF percolated phase: (a) $p = 0.01$, (b) $p = 0.1$; The ferromagnetic phase: (c) $p = 0.3$, (d) $p = 0.9$. Black sites are filled.*

Thus we need to find another order parameter to more adequately describe the structural changes in the region identified by the mean field as the (F+AF)-phase. To do this we note that at $p = 0$ the regions with the sublattice disbalance $\lambda = \varphi_A - \varphi_B \neq 0$ are formed by the directed percolation process in the time direction, cf. Fig.1. So it is natural to assume that at least some remnants of such process survive for $p \neq 0$. Indeed, we find by inspection of the various steady states patterns that while the AF domains always have a finite life time at $p \neq 0$, at small $p$ two types of those domains are however very often in the immediate contact with each other, separated by the states 00 or 11, where zeros and unities belong to different domains, see Fig.7. Then our conjecture is that at small $p$ there exist connected self-avoiding paths through the whole lattice in the time direction such that the path either lies within a single AF domain, or hopping between different neighboring domains across 00 or 11 pairs is allowed without breaking the continuity of the path. The existence of so defined directed percolation process can be a hallmark of the new phase which is predicted as the (F+AF)-phase by the mean-filed approximation. Then the capacity of percolation cluster can be identified as the order parameter for the transition into this phase.

To numerically verify this conjecture we introduce local AF parameters defined on the pairs of the neighboring spatial sites:

$$\lambda_{2j,2\tau} = n_{2j+1,2\tau} - n_{2j,2\tau}$$
$$\lambda_{2j+1,2\tau+1} = n_{2j+2,2\tau+1} - n_{2j+1,2\tau+1} \qquad (21)$$
$$j = 0,1...,N/2-1, \quad \tau = 0,1...,T/2-1$$

By breaking the original space-time lattice into pairs of sites in agreement with the above equations, one can introduce a dual lattice (rotated on the angle $\pi/4$ with respect to the original $2d$ lattice) where the local variables $\lambda_{i,t}$ reside. This is shown in Fig.8a. Since there are two possible degenerate AF ordering patterns, the cells with $\lambda_{i,t}=1$ and $\lambda_{i,t}=-1$ belong to different AF clusters. The latter correspond to the occupied sublattice A or the sublattice B, respectively, cf. Fig.1(a).

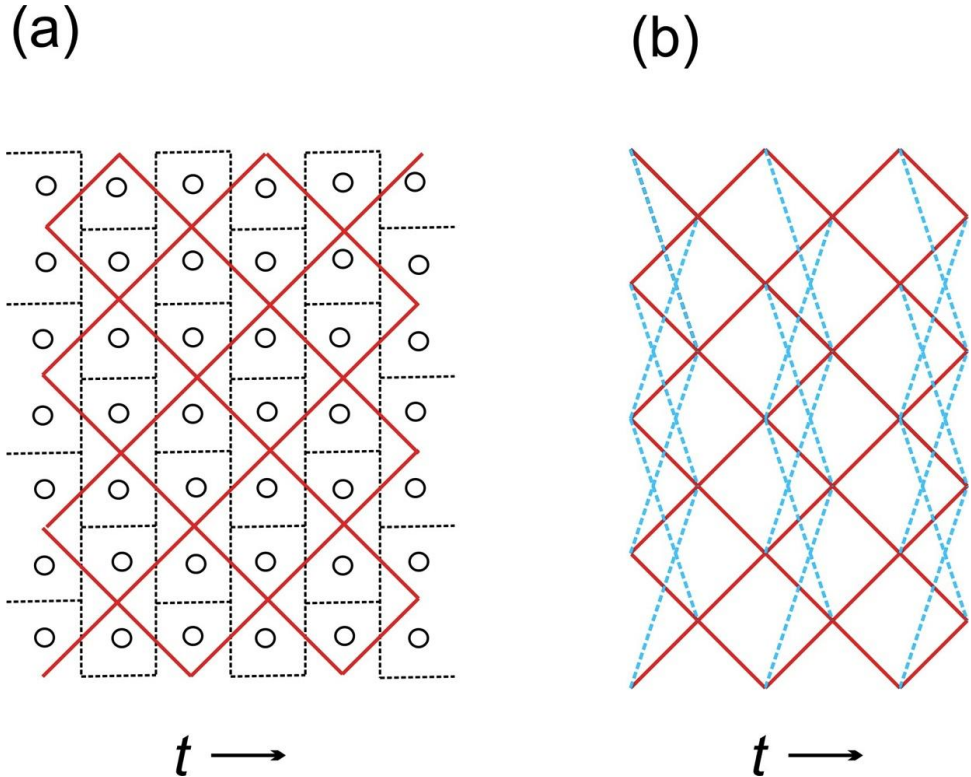

Fig.8. *(Color online) (a) Separation of the original lattice onto the two-site cells (dotted lines) and the dual square lattice where the local AF variables λ reside (solid lines); (b) decorated dual lattice with additional bonds (dashed lines) providing the percolation between different AF clusters.*

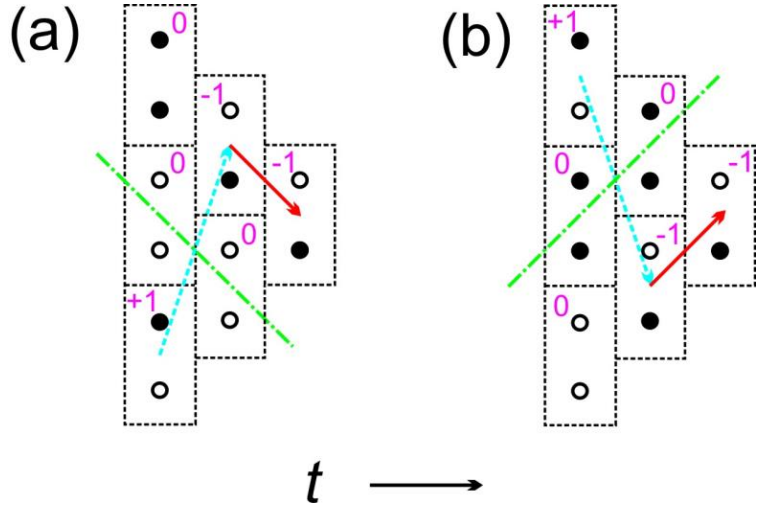

Fig.9. *(Color online) Percolation between different contacting AF clusters. Black circles are filled sites. Dash-dotted lines separate clusters; directed solid arrows indicate the nearest hoppings (bonds of the dual lattice in Fig. 8a) which provide percolation within AF clusters ; directed dotted arrows indicate the next-nearest hoppings (additional bonds of the decorated dual lattice in Fig. 8(b)) which provide percolation between AF clusters. The numbers shown inside the cells are the values of $\lambda$.*

We analyze the percolation on the lattice in the time direction. The percolation is allowed between the nearest or next-nearest neighboring sites of the dual lattice with $|\lambda_{i,t}|=1$, i.e. the sites belonging to the both types of the AF clusters. These percolating bonds of the dual lattice are shown in Fig. 8(b). The additional next-nearest neighboring dashed bonds are introduced to make possible the percolation between two different AF clusters separated by a pair of 00 or 11 sites. We show in Fig. 9 the allowed percolation steps between the nearest and next-nearest neighboring sites in terms of the original lattice.

We have studied the possibility of this type of directed percolation in the steady states patterns at small *p*. We have found that such phase exists indeed in the region roughly corresponding to the mean-field F+AF phase. The examples of such "AF percolation" patterns are shown in Fig. 10. This new AF percolation phase is presented in the exact numerical phase diagram of the model presented in Fig. 4. Several critical values of $p_c(q)$ where the percolation vanishes are given in Table 3.

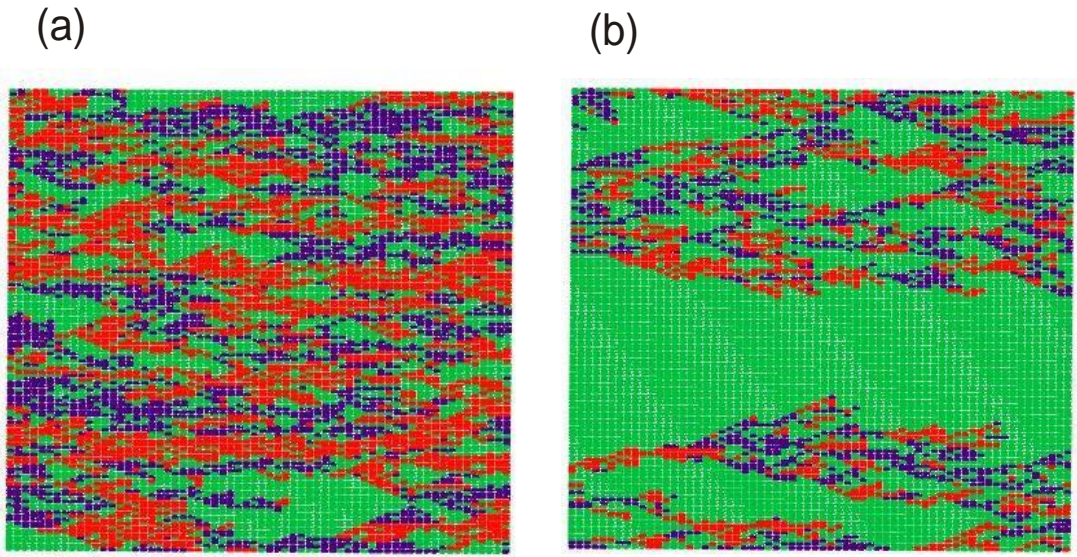

Fig. 10. *(Color online) Percolation patterns for q = 0.9, p = 0.15 (a) and q = 0.9, p = 0.22 (b). Red and blue squares are percolating AF cells with $\lambda = 1$ and $\lambda = -1$, respectively. The green background stands either the for the non-AF cells with $\lambda = 0$, or for the disconnected AF cells.*

To detect the AF percolation we proceed as follows: we compute $\lambda_{i,t}$ for two-site cells in a given 2$d$ evolution pattern and then enumerate first the AF cells with $|\lambda_{i,t}| = 1$ at $t = 2$ which can be reached from the AF cells at $t = 0$ by hopping over the bonds presented in Fig. 8(b). Then the similar hopping from the obtained percolating set of cells at $t = 2$ to the AF cells at $t = 4$ are considered and so on. Thus at every even time step we get a number of cells which can be reached from the initial state through the directed paths on the decorated square lattice shown in Fig. 8(b).

Thus we determine the time dependencies of the mean fraction of the percolating cells $\mathcal{P}(t)$ (at even time steps) for samples with $N = 10^5$ sites for 2000 time steps by averaging over 1000 trials as was done before for A-F transition. The example of using these data for determination of critical point, $\alpha$ and $\nu_{\parallel}$ is given in Fig. 11 (a, b). Similarly to find index $z$ we perform the finite-size scaling analysis of critical relaxation of $\mathcal{P}(t)$ in samples with $N$ = 100, 200, 400 sites for $T = 5\cdot10^3$, $1.5\cdot10^3$, $5\cdot10^4$ time steps averaging data over $2\cdot10^4$, $10^4$ and $5\cdot10^3$ trials correspondingly, see Fig. 11(c). The results are shown in Table 3. Again, since we obtain the values of the indices very close to the DP ones, we conclude that the hidden F-AF transition belongs to that universality class. Thus we have obtained unambiguous evidences of existence of the two active phases in our model and of the second order phase transition between them.

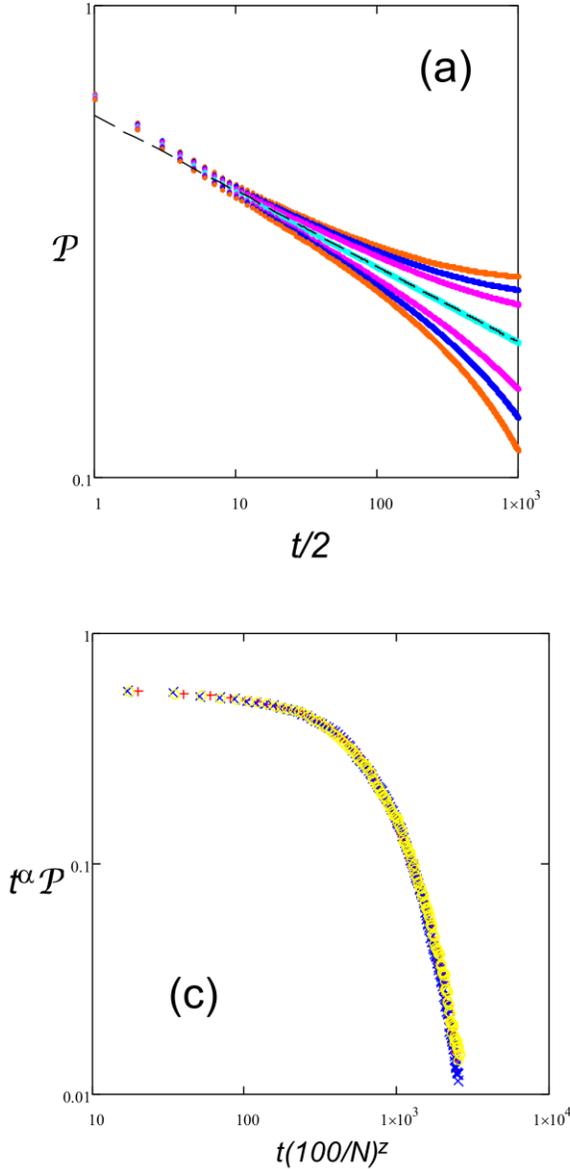

Fig.11. *(Color online) (a) MC simulations of $\mathcal{P}(t)$ relaxation for q = 0.8 and series of p near $p_c \approx 0.218$, from top to bottom: p=0.21, 0.212, 0.214, 0.218, 0.222, 0.224, 0.226. Dashed line corresponds to power law $0.58/t^\alpha$, with $\alpha = 0.16$. (b) Collapse of the curves from (a) onto a single scaling function. Fitting gives the values of $\nu_{||}$ and $p_c$ shown in Table 3. (c) Relaxation of the AF order parameter $\mathcal{P}(t)$ for chains with numbers of sites N =100 (x), N =200 (+), N =400 (o) at $p_c$ collapsed onto a single scaling function yields the index z.*

Table 3. *Critical points $p_c$ and critical indices for several values of q for the AF percolation transition.*

| q | $p_c$ | $\alpha$ | $\nu_{||}$ | z | $\beta=\alpha\nu_{||}$ | $\nu_\perp=\nu_{||}/z$ |
|---|---|---|---|---|---|---|
| 0.7 | 0.1258(2) | 0.16(01) | 1.75(7) | 1.56(5) | 0.28(1) | 1.12(3) |
| 0.8 | 0.2181(2) | 0.16(01) | 1.74(7) | 1.55(8) | 0.28(01) | 1.12(1) |
| 0.9 | 0.2364(65) | 0.16(01) | 1.74(9) | 1.58(2) | 0.28(01) | 1.10(6) |
| 1 | 0.2274(5) | 0.16(01) | 1.74(2) | 1.60(5) | 0.27(9) | 1.08(5 |

Another independent sign of the phase transition is provided by the behavior of the average length of the spatial AF domains $L$:

$$L = \left[\rho(1,1)+\rho(0,0)\right]^{-1} = \left[1-\rho(1,0)-\rho(0,1)\right]^{-1} \equiv (1-n_{10})^{-1}$$

where $n_{10} = N^{-1}\sum_i \delta(n_i + n_{i+1}, 1)$ is determined from the MC data in the stationary states. The interpretation of $L$ as the spatial domain length is valid for $q$ close to 1 and $p \ll 1$ when almost all pairs 00 and 11 correspond to the boundaries between AF domains. So $\rho(1,1) + \rho(0,0)$ is the domain wall density and its inverse is the average domain length. Apparently meaningful values of $L$ must be greater than 2 and this is true for small $p$. As one can see from the numerical results shown in Fig. 12, $L$ starts to deviate from the linear dependence (empirical fit) $L = 3 - 2p$ at $L \approx 2.5$. The values of $p$ when this deviation becomes noticeable corresponds roughly to the transition points in Table 3 for $q = 0.8, 0.9, 1$. So the AF percolation phase in Fig. 4 where $L > 2.5$ has an extensive number of the spatial AF clusters in agreement with the physical meaning of the percolating AF order.

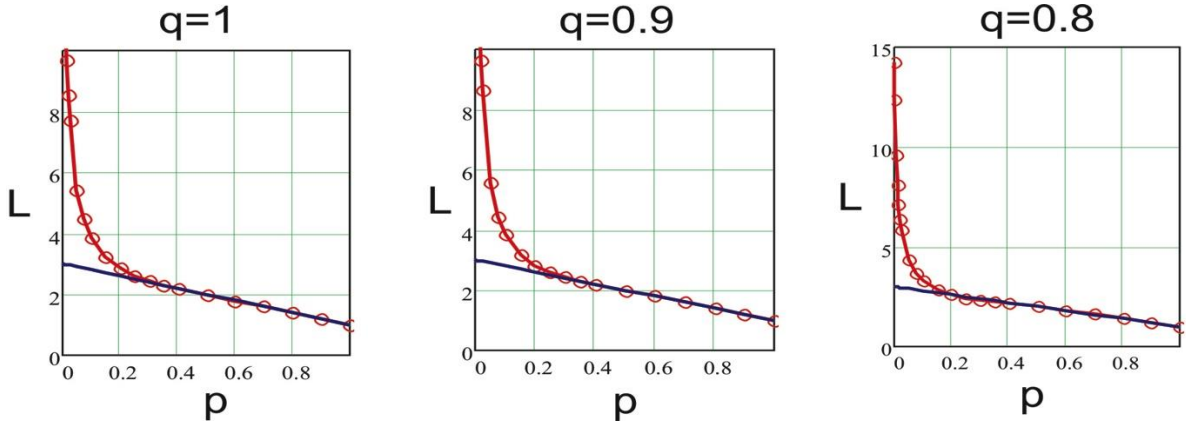

Fig.12. *(Color online) p – dependence of L for q = 1, 0.9, 0.8. Straight lines show the linear fit L=3-2p.*

### 5. Statistical mechanics representation of the PCA.

To relate the present kinetic theory to statistical mechanics we would like to represent $W(\mathbf{n}_T | \mathbf{n}_0)$ in Eq. (2) as a partition function of a certain Gibbs distribution. However we cannot put $\prod_{t=0}^{T-1} W(\mathbf{n}_{t+1} | \mathbf{n}_t) = e^{-H} > 0$ directly in (2) for the arbitrary $n_{t,i} = 0, 1$ since some $W(\mathbf{n}_{t+1} | \mathbf{n}_t) = 0$ because of the condition $W_{i,t}(1|0,0,0) = 0$. Still we can make such relation on a restricted phase space putting

$$\prod_{t=0}^{T-1} W(\mathbf{n}_{t+1} | \mathbf{n}_t) = \delta(V, 0) e^{-H_0}$$

Here

$$V = \sum_{t,i} V_{t,i} \ , \quad V_{t,i}\left(n_{t+1,i}|n_{t,i-1},n_{t,i},n_{t,i+1}\right) = n_{t+1,i}\left(1-n_{t,i}\right)\left(1-n_{t,i-1}\right)\left(1-n_{t,i+1}\right)$$

and the Kroneker delta which is zero if at least one term $V_{i,t}(1|0,0,0)=1$ appears, thus implementing the condition $W_{i,t}(1|0,0,0)=0$ for all $i$ and $t$. We can now define the Hamiltonian $H_0$ on the rest of the phase space where the configurations $(1|0,0,0)$ are excluded, getting

$$H_0 = \sum_{t,i} H_{t,i} \ , \quad H_{t,i} = -\ln W_{t,i} = -n_{t+1,i}\ln\left(P_{t,i}\right) - \left(1-n_{t+1,i}\right)\ln\left(1-P_{t,i}\right)$$

Using the representation

$$\delta(V,0) = \int_{-\pi}^{\pi} \frac{d\psi}{2\pi} e^{-i\psi V}$$

we finally obtain

$$W\left(\mathbf{n}_T|\mathbf{n}_0\right) = \sum_{\mathbf{n}_1,\ldots,\mathbf{n}_{T-1}} \int_{-\pi}^{\pi} \frac{d\psi}{2\pi} e^{-H(\psi)}$$

$$H(\psi) = \sum_{t,i}\left[-n_{t,i}\ln(1-p)(1-q)^2 + n_{t+1,i}n_{t,i}\ln\frac{1-p}{p} + 2n_{t,i}n_{t,i+1}\ln(1-q)\right]$$

$$+ \sum_{t,i} n_{t+1,i}\left(1-n_{t,i}\right)\left[\left(n_{t,i-1}+n_{t,i+1}\right)\ln\frac{1-q}{q} + n_{t,i-1}n_{t,i+1}\ln\frac{q}{2-q} + \left(1-n_{t,i-1}\right)\left(1-n_{t,i+1}\right)i\psi\right] \quad (22)$$

Thus we have established the equivalence of the $1d$ kinetic model to the statistical mechanics of the $2d$ lattice gas with the restricted phase space, which is achieved by introducing an additional degree of freedom $\psi$.

This formulation of the present kinetic process can be used to study the critical properties of, e.g., the F – A transition with a quite simple order parameter $\rho = \sum_{i=1}^{N} n_i / N$ using the renormalization group methods. One can note that the condition $V_{i,t}(1|0,0,0) = 0$ allows us to eliminate the fourth order term over fluctuations of $\rho$ in Eq. (22) (apart from the term proportional to $\psi$). As the short-range restriction of the phase space seems to be irrelevant for the F – A transition, so its critical properties will be determined by the third order interactions, the same as in the Kinzel-Domany automaton [3] belonging to the DP universality class. This explains our Monte Carlo results on the critical indices of that transition.

As we have shown the model with the Hamiltonian (22) undergoes also a transition between the ferromagnetic and the AF percolation phases characterized by a non-local order parameter, namely, the capacity of the directed percolation cluster composed of AF cells. Since the work of Kinzel [3] it is known that Hamiltonians with short-range interactions can describe such phase transitions but it does not seem to be realized even by now that the percolation order parameter can be recast in the from very similar to that of the well known nonlocal string order parameters introduced first for some quantum spin models [15]. The DP order parameter on the simple square lattice of Fig. 8(a) can be introduced using the variable defined as a product (string) of the time-ordered occupation numbers $v_{i,t}$:

$$O_{i,T}(\sigma) = v_{i,T} \prod_{t=1}^{T-1} v_{i+\sum_{k=t}^{T-1} \sigma_k, t} \qquad (23)$$

Here all parameters $\sigma$ can admit two values $\sigma_k = \pm 1$ ($k = 1,...,T-1$). It is easy to understand by writing the r.h.s. of Eq. (23) starting from the last term backward in time, that each particular set of the values $\sigma_k$ corresponds to a time-directed path starting from some spatial point at $t = 0$ and ending at the $i$-th site at $t = T$. The string variable $O_{i,T}(\sigma)$ is 1 if a given path is percolating, i.e., all $v_{i,t} = 1$ along this path, and zero otherwise. The averaging of (23) with the effective Gibbs distribution (22) and summation over all possible directed paths gives the probability that at $t = T$ the $i$-th site belongs to the percolation cluster:

$$\mathcal{P}_{i,T} = \sum_{\sigma_1,...\sigma_{T-1}} \langle O_{i,T}(\sigma) \rangle \qquad (24)$$

Actually in spatial ring geometry $\mathcal{P}_{i,T}$ does not depend on site index, so in the thermodynamic limit $T, N \to \infty$ Eq. (24) yields the DP order parameter.

To define the AF percolation order parameter in our model we need to use $v_{i,t} = |\lambda_{i,t}|$ (cf. Eq. (21)) in the definition (23). Also for the decorated square lattice (see Fig. 8(b)) we need $\sigma_{2k-1} = \pm 1, \pm 2$, $\sigma_{2k} = \pm 1$ to parameterize the path. Then the thermodynamic limit in Eq. (24) with the so defined string operator yields the AF percolation order parameter $\mathcal{P}$. In principle, such procedure can be generalized for arbitrary lattice to calculate the percolation order parameter via the string operator.

The string variable $O_{j,t}(\sigma)$ can also be related to the commonly used *pair connectedness function* $C_{i,0;j,t}$ [1] which is the probability that the site $i$ at the time $t=0$ is connected to the site $j$ at the time $t$ via a percolating path. The relation is:

$$C_{i,0;j,t} = C_{i-j,t} = \sum_{\sigma_1,\sigma_2,...\sigma_{t-1}} \langle O_{j,t}(\sigma) \rangle \delta_{i,j+\sum_{k=1}^{t-1} \sigma_k}$$

The dependence of $C_{i,0;j,t}$ on the difference $i$-$j$ follows from the translational invariance in the spatial ring geometry. This function can be used for analytical and numerical computations of other critical indices of the model [1, 16].

## 6. Conclusions

In the present paper the one-dimensional kinetic contact process with parallel update is introduced and studied by the mean-field approximation and Monte Carlo simulations. Contrary to a more conventional two-phase scenario for these types of models with single active and absorbing phases, the special property of the present kinetic process is the phase diagram with two different adjacent active phases separated by a line of the second-order transition, and a tricritical point where the active and the absorbing phases meet. The phase diagram of the proposed model constitutes the main result of this study. While one of the active phases is quite standard with a smooth average filling of the space-time lattice, the second active phase demonstrates a very subtle (hidden) percolating order. The transition is hidden in the sense that it cannot be detected via a simple visual inspection of the steady states patterns as in other cases known to us. Qualitatively, the nature of this phase is related to existence of a strong short-range order which appears through the multitude of finite antiferromagnetic clusters on the space-time lattice. Contrary to a conventional antiferromagnetic phase where the long-range order is due to the critical growth of those clusters, in our model the appearance of the new phase is due not to the growth but to the proliferation of the both types of antiferromagnetic clusters, resulting in eventually the possibility to percolate through those clusters across the whole lattice in the time direction. To deal with this order quantitatively, either numerically or analytically, we proposed to change the local variables and map the model onto a dual lattice. We dubbed this process the (hidden) antiferromagnetic percolation. We find the numerical values of the critical indices via scaling analysis for the both transitions (F-A and F-AF percolation) which strongly suggest that they belong to the DP universality class.

We also give a new (to best of our knowledge) definition of the percolation order parameter in terms a nonlocal (string) operator. We used this definition to calculate the order parameter on the dual lattice in the antiferromagnetic percolating phase, but the definition is valid for the standard DP as well, and can be straightforwardly generalized for arbitrary lattice. The proposed definition of the percolation order parameter as a result of the averaging over possible string variables (paths) points out its relation to the similar nonlocal string order parameters known from studies of the low-dimensional magnetic, fermionic or bosonic systems with hidden orders where the conventional Landau order is absent [13,15-17]. Thus we can suggest that the similar nonlocal string operators can be used to establish hidden order parameters in other $1d$ kinetic models [1, 2].

In the context of related recent work [6-10] on additional active and/or absorbing phases in similar models, we would like to single out the model considered by de Oliveira and Dickman [8]. Their model in $d \geq 2$ demonstrates the second active phase analogous to our mean-field F+AF phase considered in Sec. 3. The model [8] in $d = 1$ seems to be a very good candidate to check for existence of the percolating phase with hidden order similar to our findings. More broadly, the present results imply that other $1d$ models of contact processes known from the literature [1,2] might possess active phases with hidden orders overlooked so far.

## Acknowledgements

G.Y.C. acknowledges support from the Laurentian University Research Fund (LURF). The numerical part of this work was made possible by the facilities of the Shared Hierarchical Academic Research Computing Network (SHARCNET) and Compute/Calcul Canada.